\documentclass[sigconf,natbib=true,screen]{acmart}
\usepackage{booktabs}
\usepackage{amsmath}
\usepackage{algorithm}
\usepackage{algpseudocode}
\setcopyright{none}
\usepackage{colortbl}
\usepackage{xcolor}
\definecolor{impcolor}{RGB}{34,139,34}
\newcommand{\impcell}[1]{\cellcolor{impcolor!#1}}

\copyrightyear{2026}
\acmYear{2026}

\setcopyright{cc}
\setcctype{by}

\acmConference[CIKM '26]
{Proceedings of the 35th ACM International Conference on Information and Knowledge Management}
{November 7--11, 2026}
{Rome, Italy}

\acmBooktitle{Proceedings of the 35th ACM International Conference on Information and Knowledge Management (CIKM '26), November 7--11, 2026, Rome, Italy}

\begin{document}

\title{Difficulty-Gated Fusion of Reasoning Views for Temporal Retrieval}
% \titlenote{Accepted as a short paper at the 35th ACM International Conference on Information and Knowledge Management (CIKM '26).}

\author{Jamie Holdcroft}
\authornote{Jamie Holdcroft and Abdelrahman Abdallah contributed equally to this work.}
\affiliation{%
  \institution{UNSW Sydney}
  \city{Sydney}
  \country{Australia}
}
\email{j.holdcroft@student.unsw.edu.au}

\author{Abdelrahman Abdallah}
\authornotemark[1]
\affiliation{%
  \institution{University of Innsbruck}
  \city{Innsbruck}
  \country{Austria}
}
\email{abdelrahman.abdallah@uibk.ac.at}

\author{Adam Jatowt}
\affiliation{%
  \institution{University of Innsbruck}
  \city{Innsbruck }
  \country{Austria}}
\email{adam.jatowt@uibk.ac.at}



\begin{abstract}
Reasoning-intensive temporal retrieval requires matching a query to documents whose relevance depends on shared temporal reasoning rather than lexical overlap. Expanding a query into several reformulations that make its temporal intent explicit, and retrieving with each, supplies this reasoning, but fusing the resulting rankings with equal weights wastes accuracy: for any single query, only some reformulations are reliable. We propose query-difficulty-gated fusion of reasoning views. From each view we read an eight-dimensional signature of its score distribution, built from query-performance-prediction quantities such as softmax entropy, score gaps, and dispersion, and a gate of roughly one thousand parameters maps these signatures to per-query view weights. The fused ranking uses no relevance labels at inference, no re-ranking, and no fine-tuning of the retriever; the gate is trained leave-one-task-out. On the \textsc{Tempo} benchmark, the method improves all six retrievers we evaluate, from BERT encoders to 7B decoder retrievers, with the largest gains on the weaker backbones. The strongest retrievers reach $0.297$ and $0.303$ nDCG@10, and the per-query gain over the original query is significant under a paired bootstrap ($p<0.001$). A per-query oracle reaches $0.364$ against our realized $0.297$, exposing headroom that identifies per-query view selection as a concrete next step.
\end{abstract}


\maketitle

\begin{figure*}[t]
  \centering
  \includegraphics[width=0.7\textwidth]{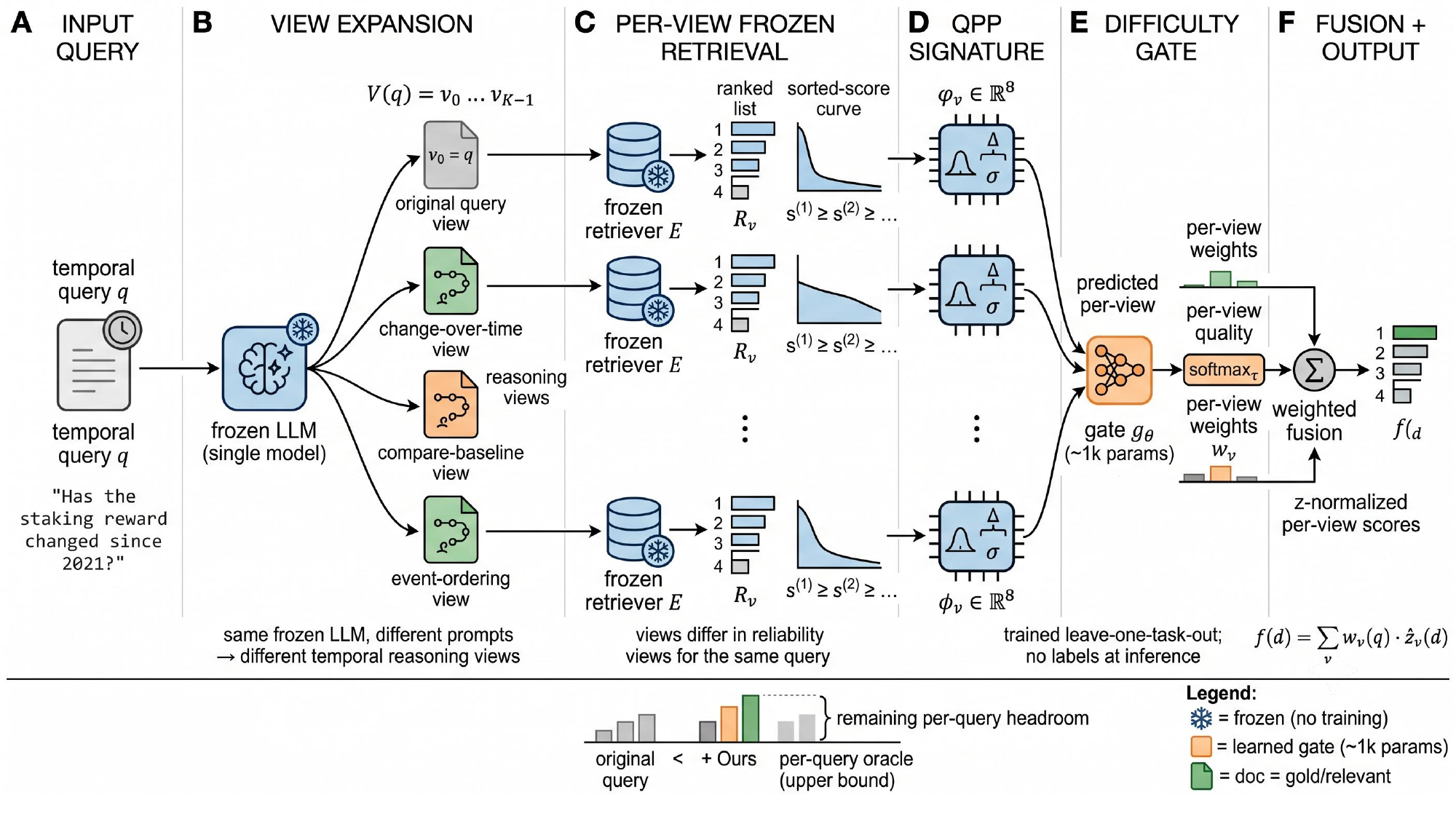}
  \caption{\textbf{Difficulty-gated fusion of reasoning views.} A temporal query is expanded into $K$ reasoning views; each is retrieved by a \emph{same frozen} retriever, yielding a ranked list and a score distribution whose shape reveals how reliable that view is for this query. A query-performance-prediction signature $\phi_v$ summarizes each view's score geometry, and a tiny gate $g_\theta$ (the only learned component, ${\sim}1$k parameters, trained leave-one-task-out) maps these signatures to per-query view weights $w_v$ that drive the fused ranking.}
  \label{fig:overview}
\end{figure*}

\section{Introduction}
\label{sec:intro}

Most retrieval methods~\citep{robertson2009bm25,karpukhin2020dense} are trained to reward lexical or topical overlap between a query and a document. Reasoning-intensive retrieval~\citep{diver} breaks this assumption: the relevance of a document is not visible on the surface and only becomes clear after several steps of inference about what the query is really asking \citep{abdallah2026llm,bright,reasonir}. The \textsc{Tempo} benchmark \citep{tempo} extends this setting along the temporal dimension. A query must be answered with evidence that is correctly grounded in time. This often means synthesizing across several periods, for example to track a change, to compare an earlier baseline against a later state, or to order events in a sequence. The combination of latent reasoning and temporal grounding makes \textsc{Tempo} dataset more challenging than lookup-style temporal question answering, and published systems have stalled \citep{tan2023tempreason,piryani2025s,abdallah2026tempretriever,chen2021timeqa}. The strongest reported model, DiVeR \citep{diver}, reaches $0.297$ nDCG@10, and a frontier reasoning retriever queried with the plain query does no better. 
% Expanding a query into several reformulations that spell out its temporal intent, and retrieving with each under a single frozen retriever, enables temporal reasoning that a plain query omits \citep{wang2023query2doc,gao2023hyde}. Different reformulations surface different facets of the information need, so pooling their shortlists raises the chance that the relevant document appears among the candidates. The central challenge, however, lies in the combination step. Equal weighting averages one reliable view against several confident but wrong ones, diluting the gain from reformulation \citep{benhamRR,benhamQV}. What is needed is a per-query estimate of each view's reliability, obtained without relevance labels at retrieval time.
A query can be expanded into several reformulations that make its temporal intent explicit; we call each reformulation, together with the original query, a \emph{view} of the information need. Retrieving with each view under a single frozen retriever enables temporal reasoning that a plain query omits \citep{wang2023query2doc,gao2023hyde}. Because different views surface different facets of the need, pooling their shortlists raises the chance that the relevant document appears among the candidates. The central challenge, however, lies in the combination step. Equal weighting averages one reliable view against several confident but wrong ones, diluting the gain from reformulation \citep{benhamRR,benhamQV}. What is needed is a per-query estimate of each view's reliability, obtained without relevance labels at retrieval time.
We obtain that judgment from query performance prediction (QPP) \citep{qppsurvey,clarity,shtok2012nqc}. For each view we read the shape of its own retrieval-score distribution and summarize it as an eight-dimensional signature $\phi_v$ that captures how peaked and well-separated the ranking is. A peaked, well-separated distribution is the fingerprint of a confident, well-posed view, and a flat one signals uncertainty. A small gate $g_\theta$ of about one thousand parameters maps each signature to a predicted view quality, and a temperature softmax across views turns these predictions into per-query weights that drive a single fused ranking. The gate is the only learned component. It is trained leave-one-task-out, so no relevance label of the evaluated task is seen at inference. There is no re-ranking and no fine-tuning of the retriever. Figure~\ref{fig:overview} shows the pipeline. Because the gate reads only score distributions, it can be attached to any retriever without modification. On \textsc{Tempo} the method improves every retriever we test, and the gain is largest for the weakest backbones and shrinks as backbone strength rises. If the improvement came from a particular embedding model, we would not expect this trend; it instead points to the fusion procedure as the source. We also measure the per-query headroom that reasoning reformulations create and how much of it the gate recovers, which frames per-query view selection as the natural next target.

Our contributions are the following. \textbf{(1)}~A retriever-agnostic, training-light fusion method that gates LLM reasoning reformulations by a per-view, per-query QPP signature read only from each view's score distribution. The gate is the sole learned component (${\sim}1$k parameters), and the method uses no re-ranking and no fine-tuning of the retriever. \textbf{(2)}~Experimental results showing that the gate's per-query weights carry usable reliability signal: difficulty-gated fusion improves every retriever we test on \textsc{Tempo} under one fixed protocol, outperforms uniform fusion, and yields a statistically significant per-query gain, with the largest improvements on the weakest backbones, which isolates the fusion procedure rather than any single retriever as the source (Section~\ref{sec:experiments}, Table~\ref{tab:main}). \textbf{(3)}~A headroom analysis that quantifies the per-query signal reasoning reformulations expose and how much of it the gate realizes, framing per-query view selection as a concrete and well-posed direction for future work.

\section{Method}
\label{sec:method}
Given a corpus $D$ and a temporal query $q$, we expand $q$ into several LLM reasoning views, retrieve each independently with a retriever, and fuse the rankings with per-query weights set by a difficulty gate. Sections~\ref{ssec:views} to \ref{ssec:train} develop this fusion; Section~\ref{ssec:oracle} defines the per-query headroom it leaves on the table.

\subsection{Reasoning views and per-view scoring}
\label{ssec:views}
% We expand $q$ into a set of $K$ reasoning views
% \begin{equation}
% V(q)=\{v_0,v_1,\dots,v_{K-1}\},\qquad v_0=q,
% \end{equation}
% where $v_0$ is the original query and $v_1,\dots,v_{K-1}$ are reasoning reformulations produced by a frozen LLM 
We expand $q$ into a set of $K$ reasoning views $V(q)=\{v_0,v_1,\dots,v_{K-1}\}$, where $v_0=q$ is the original query and $v_1,\dots,v_{K-1}$ are reasoning reformulations produced once offline by GPT-4o \citep{openai2024gpt4o}. Each reformulation targets a distinct mode of temporal reasoning, mirroring the temporal phenomena \textsc{Tempo} is built around: a \emph{change-over-time} view rewrites $q$ to track how a quantity evolves across periods, a \emph{compare-baseline} view contrasts an earlier state against a later one, and an \emph{event-ordering} view makes the temporal sequence of events explicit (Figure~\ref{fig:overview}, panel~B). These views surface complementary facets of the temporal information need, while the retriever itself remains frozen.
%that make the query's temporal intent and latent inference steps explicit. 
Let $E_r$ be the frozen encoder of retriever $r$, with its own native pooling and its own query and document instruction prefixes $\rho^q_r,\rho^d_r$. Each view is scored independently against $D$ by inner product in retriever $r$'s embedding space,
\begin{equation}
s_v(d)=\big\langle\, E_r(\rho^q_r \oplus v),\; E_r(\rho^d_r \oplus d)\,\big\rangle,\qquad d\in D,
\label{eq:score}
\end{equation}
where $\oplus$ denotes concatenation. The rest of the method consumes only the resulting score lists, so it is independent of how $E_r$ is built. Let $R_v=\mathrm{topN}(s_v)$ be the $N$ highest-scoring documents under view $v$. 
% The candidate pool for $q$ is the union of all views' shortlists,
% \begin{equation}
% C(q)=\bigcup_{v\in V(q)} R_v .
% \label{eq:union}
% \end{equation}
% Different reasoning views surface different facets
The candidate pool for $q$ is the union of all views' shortlists, $C(q)=\bigcup_{v\in V(q)} R_v$. Different reasoning views surface different facets
of the temporal information need, so the union $C(q)$ pools recall across views and raises the ceiling on what any downstream weighting can recover.

\subsection{Per-view z-normalization}
View scores are not directly comparable, because each $v$ induces its own score scale, and naively summing $s_v(d)$ would allow a single high-variance view to dominate the fused score. We stabilize each view by per-query z-normalization over its own top-$N$ scores, with mean $\mu_v$ and standard deviation $\sigma_v$ of $\{s_v(d):d\in R_v\}$:
\begin{equation}
\hat{z}_v(d)=
\begin{cases}
\dfrac{s_v(d)-\mu_v}{\sigma_v}, & d\in R_v,\\[1.2ex]
\min_{d'\in R_v}\hat{z}_v(d')-1, & d\notin R_v .
\end{cases}
\label{eq:znorm}
\end{equation}
The floor, one standard deviation below view $v$'s weakest retained candidate, assigns a controlled penalty to documents that view $v$ never retrieved, so a document found by one view but missing from another is not silently treated as average. After Eq.~\eqref{eq:znorm} every view contributes on a common, zero-mean, unit-variance scale.
\begin{table*}[t]
\centering
\caption{Reasoning-view fusion gated by query difficulty (QPP) on all 13 \textsc{Tempo} tasks (nDCG@10, $\times100$, $K{=}3$ views).
% Every retriever is reproduced with the benchmark's official code on raw text; for each we report the original-query baseline and the result after adding our method (\textbf{+ Ours}). The method uses no fine-tuning of the retriever and no re-ranking stage. It improves all six retrievers, with the largest gains on the weaker backbones, indicating that the benefit comes from the difficulty-gated fusion rather than from a particular embedding model. Cell shading on the \textbf{+ Ours} rows is proportional to the per-task improvement over the corresponding baseline.
}
\label{tab:main}
\setlength{\tabcolsep}{4pt}
\small
\resizebox{0.8\textwidth}{!}{%
\begin{tabular}{lcccccccccccccc}
\toprule
Retriever & Bitcoin & Cardano & Iota & Monero & Econ. & Law & Pol. & Quant & Travel & Work. & Geneal. & HSM & Hist. & \textbf{Mean} \\
\midrule
Contriever & 12.5 & 12.2 & 38.1 & 9.8 & 15.0 & 22.3 & 28.7 & 10.7 & 20.8 & 20.6 & 24.5 & 17.2 & 22.7 & 19.6 \\
\quad + Ours & 12.3 & \impcell{55}\textbf{21.3} & 36.9 & \impcell{35}\textbf{14.1} & \impcell{50}\textbf{21.5} & \impcell{55}\textbf{30.3} & \impcell{50}\textbf{35.6} & \impcell{40}\textbf{15.9} & \impcell{12}\textbf{22.3} & \impcell{90}\textbf{40.8} & \impcell{60}\textbf{32.8} & \impcell{60}\textbf{25.9} & \impcell{38}\textbf{27.4} & \impcell{48}\textbf{25.9} \\
\midrule
BGE & 13.2 & 12.4 & 36.1 & 14.1 & 11.1 & 26.7 & 25.5 & 10.6 & 19.6 & 24.6 & 21.5 & 20.6 & 23.3 & 19.9 \\
\quad + Ours & \impcell{25}\textbf{16.3} & \impcell{60}\textbf{21.0} & \impcell{8}\textbf{37.1} & \impcell{35}\textbf{18.7} & \impcell{30}\textbf{15.0} & \impcell{35}\textbf{31.2} & \impcell{42}\textbf{30.8} & \impcell{6}\textbf{11.3} & \impcell{12}\textbf{21.0} & \impcell{75}\textbf{34.9} & \impcell{16}\textbf{23.5} & \impcell{52}\textbf{27.4} & \impcell{16}\textbf{25.3} & \impcell{34}\textbf{24.1} \\
\midrule
Inst-L & 13.5 & 13.7 & 34.3 & 16.1 & 15.9 & 31.7 & 29.1 & 13.7 & 21.6 & 30.3 & 23.8 & 21.8 & 24.0 & 22.3 \\
\quad + Ours & 13.4 & \impcell{28}\textbf{17.1} & \impcell{70}\textbf{43.4} & 13.7 & \impcell{38}\textbf{20.5} & 25.6 & \impcell{62}\textbf{37.0} & 9.4 & 18.9 & \impcell{58}\textbf{37.9} & \impcell{45}\textbf{29.2} & \impcell{68}\textbf{30.2} & 23.3 & \impcell{20}\textbf{24.6} \\
\midrule
SBERT & 13.0 & 20.4 & 33.2 & 14.3 & 12.3 & 27.5 & 31.4 & 15.0 & 22.4 & 30.5 & 23.2 & 23.4 & 24.1 & 22.4 \\
\quad + Ours & \impcell{6}\textbf{13.7} & \impcell{10}\textbf{21.7} & 32.6 & \impcell{18}\textbf{16.4} & \impcell{24}\textbf{15.2} & \impcell{45}\textbf{33.2} & \impcell{20}\textbf{33.9} & \impcell{38}\textbf{19.6} & \impcell{24}\textbf{25.2} & \impcell{52}\textbf{36.7} & \impcell{4}\textbf{23.3} & \impcell{32}\textbf{27.2} & \impcell{6}\textbf{24.8} & \impcell{22}\textbf{24.9} \\
\midrule
SFR & 15.6 & 26.6 & 37.1 & 21.5 & 19.6 & 34.1 & 39.4 & 15.0 & 24.4 & 24.7 & 31.0 & 29.2 & 26.3 & 26.5 \\
\quad + Ours & \impcell{18}\textbf{17.7} & \impcell{32}\textbf{30.3} & \impcell{22}\textbf{39.7} & 19.5 & \impcell{34}\textbf{23.5} & \impcell{28}\textbf{37.4} & \impcell{30}\textbf{42.9} & \impcell{44}\textbf{20.1} & 23.3 & \impcell{100}\textbf{37.0} & \impcell{46}\textbf{36.3} & \impcell{34}\textbf{33.0} & 25.4 & \impcell{28}\textbf{29.7} \\
\midrule
E5 & 14.6 & 33.0 & 43.1 & 17.6 & 21.6 & 29.6 & 41.7 & 12.6 & 23.1 & 27.3 & 32.5 & 32.5 & 22.9 & 27.1 \\
\quad + Ours & \impcell{28}\textbf{17.8} & \impcell{42}\textbf{37.9} & \impcell{12}\textbf{44.5} & 16.2 & \impcell{22}\textbf{24.2} & \impcell{28}\textbf{32.9} & \impcell{26}\textbf{44.7} & \impcell{40}\textbf{17.3} & 19.5 & \impcell{100}\textbf{44.8} & \impcell{54}\textbf{38.8} & \impcell{16}\textbf{34.3} & 21.2 & \impcell{28}\textbf{30.3} \\
\bottomrule
\end{tabular}%
}
\end{table*}
\subsection{QPP signature}
We summarize how confident view $v$ is for query $q$ from the shape of its sorted score sequence $s^{(1)}\!\ge\!s^{(2)}\!\ge\!\cdots$. Following score-distribution QPP \citep{clarity,shtok2012nqc}, a peaked and well-separated score distribution indicates that the view is well-posed and the retrieval task is well-defined for this query, whereas a flat distribution indicates uncertainty. The first component of the signature is a softmax entropy over the top $100$ scores,
\begin{equation}
p_i=\frac{\exp(\alpha\, s^{(i)})}{\sum_{j=1}^{100}\exp(\alpha\, s^{(j)})},\qquad
H_v=-\sum_{i=1}^{100} p_i\log p_i,
\label{eq:entropy}
\end{equation}
with temperature $\alpha=20$. The remaining seven features are the score gaps $\Delta_{1,2}=s^{(1)}-s^{(2)}$, $\Delta_{1,5}=s^{(1)}-s^{(5)}$, and $\Delta_{1,10}=s^{(1)}-s^{(10)}$, the mean and standard deviation of $s^{(1:100)}$, a decay slope $(s^{(1)}-s^{(20)})/20$, and the maximum score $s^{(1)}$:
\begin{equation}
\phi_v=\big[\,H_v,\ \Delta_{1,2},\ \Delta_{1,5},\ \Delta_{1,10},\ \overline{s^{(1:100)}},\ \mathrm{std}(s^{(1:100)}),\ \tfrac{s^{(1)}-s^{(20)}}{20},\ s^{(1)}\,\big].
\label{eq:phi}
\end{equation}
Lower entropy and larger gaps signal a sharply ranked, high-confidence view. No single coordinate of $\phi_v$ predicts reliability on its own. Section~\ref{sec:experiments} shows that the gate, reading the full signature jointly, consistently beats uniform fusion, which confirms the score-distribution geometry carries usable per-query signal.

\subsection{Difficulty gate and fused score}
A gate $g_\theta:\mathbb{R}^8\!\to\!\mathbb{R}$, a small multilayer perceptron of about one thousand parameters, maps each signature to a predicted per-view retrieval effectiveness $g_\theta(\phi_v)\approx\mathrm{nDCG@10}(R_v,q)$. We turn predictions into per-query, per-view weights with a temperature-$\tau$ softmax across views,
\begin{equation}
w_v(q)=\frac{\exp\!\big(\tau\, g_\theta(\phi_v)\big)}{\sum_{v'\in V(q)}\exp\!\big(\tau\, g_\theta(\phi_{v'})\big)},\qquad
\sum_{v} w_v(q)=1,
\label{eq:weights}
\end{equation}
and combine the normalized view scores into one fused score over the candidate pool,
\begin{equation}
f(d)=\sum_{v\in V(q)} w_v(q)\,\hat{z}_v(d),\qquad d\in C(q).
\label{eq:fused}
\end{equation}
The final ranking sorts $C(q)$ by $f(d)$ in descending order. Eq.~\eqref{eq:weights} concentrates mass on the views the gate predicts to be reliable for this query, with $\tau$ interpolating between uniform fusion at $\tau\!=\!0$ and hard routing as $\tau\!\to\!\infty$, and Eq.~\eqref{eq:fused} is a per-query convex combination of the comparable signals produced by Eq.~\eqref{eq:znorm}.

\subsection{Training objective}
\label{ssec:train}
We fit $g_\theta$ by regressing predicted onto true per-view nDCG with a mean-squared-error loss,
\begin{equation}
\mathcal{L}(\theta)=\frac{1}{|Q|}\sum_{q\in Q}\frac{1}{|V(q)|}\sum_{v\in V(q)}\Big(g_\theta(\phi_v)-\mathrm{nDCG@10}(R_v,q)\Big)^2 .
\label{eq:loss}
\end{equation}
To preclude label leakage, every evaluation uses leave-one-task-out cross-validation: for each held-out \textsc{Tempo} task the gate is trained only on the relevance labels of the other tasks, and no label of the held-out task is seen at inference. The gate therefore learns a task-agnostic mapping from score geometry to view reliability, which is what lets a single trained gate transfer across tasks and, as Section~\ref{sec:experiments} shows, across retrievers.

\subsection{Per-query headroom}
\label{ssec:oracle}
To measure how much per-query signal the views contain, we define the per-query oracle, which uses the relevance labels to pick each query's best view and is therefore an upper bound rather than a deployable method:
\begin{equation}
O=\frac{1}{|Q|}\sum_{q\in Q}\ \max_{v\in V(q)}\ \mathrm{nDCG@10}(R_v,q).
\label{eq:oracle}
\end{equation}
Writing $F$ for the realized fused effectiveness, that is, the mean nDCG@10 of the ranking induced by $f(\cdot)$ in Eq.~\eqref{eq:fused}, the quantity $O-F$ measures the headroom that an ideal per-query view selector could still recover from the same candidate pool. Adding reasoning views raises $O$, because additional views increase the probability that at least one shortlist contains the relevant document. Section~\ref{sec:experiments} shows that our gate captures a significant and consistent share of this headroom across retrievers, and quantifies the remainder as a target for future selectors.

\section{Experiments}
\label{sec:experiments}

\textbf{Setup.}
% We evaluate on the \textsc{Tempo} benchmark \citep{tempo}, which comprises 13 temporal reasoning-intensive retrieval tasks drawn from distinct StackExchange domains, and we report nDCG@10. 
We evaluate on the \textsc{Tempo} benchmark \citep{tempo}, which comprises 13 temporal reasoning-intensive retrieval tasks drawn from distinct StackExchange domains, and we report nDCG@10. 
% We focus on \textsc{Tempo} because it is, to our knowledge, the only retrieval benchmark that combines temporal grounding with reasoning-intensive relevance, and its queries are real-world questions whose answers require synthesizing dated evidence rather than matching surface terms. \textsc{Tempo} is also substantially harder than other reasoning-retrieval datasets: even strong retrievers trail their non-temporal performance by a wide margin \citep{tempo}, which leaves clear room for a method that better exploits each query's reasoning views.
We focus on \textsc{Tempo} because it is the only benchmark pairing temporal grounding with reasoning-intensive relevance over real-world queries, and is markedly harder than non-temporal reasoning-retrieval datasets \citep{tempo}, leaving clear room for better use of each query's reasoning views.
We study six retrievers that span a wide range of architectures and scales: Contriever~\citep{izacard2021unsupervised}, BGE~\citep{xiao2024c}, SBERT~\citep{reimers2019sentence}, and Instructor (Inst-L)~\citep{su2023one}, and SFR~\citep{SFRAIResearch2024} and E5~\citep{wang2022text}. To ensure a rigorous and reproducible comparison, we reproduce every retriever with the benchmark's own released code, applied to the raw document and query text and using each model's native pooling and instruction prefixes. Our reproduced baselines lie within about $0.02$ nDCG of the published averages, the usual cross-hardware reproduction tolerance. %The reasoning reformulations are generated once offline with GPT-4o \citep{openai2024gpt4o} using a fixed prompt, and the same generated views are reused for every retriever. This fixes the view set across backbones and isolates the effect of the retrieval model and the proposed difficulty-gated fusion. 
The reasoning reformulations are generated once offline with GPT-4o \citep{openai2024gpt4o} using a single fixed prompt, with no access to relevance labels, retrieved documents, or held-out task information. The same generated views are reused for every retriever, which fixes the view set across backbones and isolates the effect of the retrieval model and the proposed fusion. Because the reformulations are identical across all six retrievers, the LLM cannot explain the cross-retriever pattern in Table~\ref{tab:main}: the gain varies with backbone strength under a fixed view set, which attributes the improvement to the difficulty-gated fusion rather than to the reformulation quality. Generating the views once offline also keeps the method cheap and fully reproducible, since the LLM is never queried at retrieval time.
% The LLM is used only for query reformulation; it is not given relevance labels, retrieved documents, or held-out task information.
% Unless noted we use $K{=}3$ views (the original query plus two reformulations), which maximizes the realized score (Table~\ref{tab:ksweep}); the difficulty gate is fit by leave-one-task-out cross-validation, so the relevance labels of the evaluated task are never used at inference.

\textbf{Main result:}
Table~\ref{tab:main} is our central result. For each of the six retrievers it reports the per-task nDCG@10 of the original-query baseline and of the same retriever after adding our difficulty-gated fusion, under one fixed protocol. The method raises the thirteen-task mean for all six backbones, by between $+2.3$ and $+6.3$ points (Table~\ref{tab:main}). Two observations are noteworthy. First, the improvement is largest for the weakest backbones and smaller for the strongest, a clear inverse relationship between base accuracy and gain. This is the signature of a method whose benefit comes from the fusion itself: a weak embedder produces noisy rankings whose per-query reliability varies, exactly what difficulty-gating corrects, while a strong embedder already concentrates mass on the right documents. Second, after fusion the two strongest retrievers reach $0.297$ (SFR) and $0.303$ (E5), matching their published single-query baselines, with a further significant per-query gain, using only a thousand-parameter gate and no reformulation training. 
The per-task numbers in Table~\ref{tab:main} show the improvement is broad rather than driven by one or two domains.
% The gains concentrate on the domains where temporal reasoning is most demanding, such as workplace, genealogy, cardano, politics, and economics, and the per-task numbers in Table~\ref{tab:main} show the improvement is broad rather than driven by one or two domains.

\begin{table}[t]
\centering
\small
\caption{\textsc{Tempo} 13 task mean nDCG@10 on the SFR backbone. The oracle uses the relevance labels to pick the best view per query and is an upper bound, not a method. $\dagger$ marks a result significantly above the original query (paired bootstrap, $n{=}1730$, $p<0.001$).
% DiVeR is the strongest published single retriever, reported with its own reformulation pipeline \citep{diver}.
}
\label{tab:sota}
\begin{tabular}{lc}
\toprule
Method & nDCG@10 \\
\midrule
Original query ($K{=}1$) & $0.265$ \\
Equal-weight fusion ($K{=}5$) & $0.284$ \\
Difficulty-gated fusion ($K{=}5$) & $0.289$ \\
\textbf{Difficulty-gated fusion ($K{=}3$)} & $\mathbf{0.297}^{\dagger}$ \\
\midrule
DiVeR, best published single retriever \citep{diver} & $0.320$ \\
Per-query oracle ($K{=}5$, upper bound) & $0.364$ \\
\bottomrule
\end{tabular}
\end{table}
Table~\ref{tab:sota} analyzes SFR results in detail. Equal-weight averaging of the z-normalized views already helps ($0.265$ to $0.284$), but the difficulty gate goes further ($0.297$ at $K{=}3$), which shows the gain is not merely from pooling views but from weighting them per query by their score-distribution geometry. 
% For reference, the strongest published single retriever on \textsc{Tempo} is DiVeR at $0.320$, obtained with its own dedicated reasoning-reformulation pipeline \citep{diver}; our method reaches a comparable level by adding a thousand-parameter gate on top of an off-the-shelf embedder, with no reformulation training. 
The per-query oracle, which uses the labels to keep each query's best view and is therefore an unexploitable upper bound, sits far higher ($0.364$ at $K{=}5$), a gap we quantify in the following paragraph.

\textbf{The per-query weighting is what helps.}
The gate's edge over equal-weight fusion ($0.297$ versus $0.284$ on SFR) confirms that the score-distribution signature $\phi_v$ carries usable per-query reliability signal: views with sharply peaked, well-separated score distributions tend to be the more trustworthy ones, and the gate learns to put weight there. Because $\phi_v$ is read only from the score lists every retriever already produces, the same gate transfers across all six backbones without modification, which is what makes the improvement in Table~\ref{tab:main} so uniform.

\textbf{Effect of the number of views.}
Table~\ref{tab:ksweep} reports the realized fused score and the per-query oracle as the number of views $K$ grows on SFR. The realized score rises from $0.265$ at one view to a peak of $0.297$ at $K{=}3$, then eases to $0.289$ at $K{=}5$, while the oracle keeps climbing, from $0.348$ at $K{=}3$ to $0.364$ at $K{=}5$, because each added reformulation occasionally surfaces a query's relevant document where no other view does. 
% Two practical implications follow from this trend. First, a small view count already provides the benefit, preserving the computational efficiency of the method: each view costs one extra encoding pass and no re-ranking. Second, the widening distance between the realized score and the oracle measures how much per-query signal the reformulations still hold. 
A small view count already captures the benefit at low cost, while the widening gap between the realized score and the oracle measures how much per-query signal the reformulations still hold.
The gate converts a significant share of it into accuracy across all six retrievers, and the remainder defines a clean, well-posed target: a per-query selector that identifies the single view ranking a query's evidence first.

\begin{table}[t]
\centering
\small
\caption{Realized difficulty-gated nDCG@10 and the per-query oracle on the SFR backbone as the number of views $K$ grows. A small $K$ already captures the realizable gain, while the oracle keeps rising and measures the per-query headroom that remains.}
\label{tab:ksweep}
\begin{tabular}{lccc}
\toprule
$K$ & 1 & 3 & 5 \\
\midrule
Realized $F$ (difficulty-gated) & 0.265 & \textbf{0.297} & 0.289 \\
Per-query oracle $O$ & 0.265 & 0.348 & 0.364 \\
\bottomrule
\end{tabular}
\end{table}

\paragraph{Significance.}
On SFR, a paired bootstrap over all $n{=}1730$ queries gives a mean per-query improvement of the $K{=}3$ gate over the original query of $\Delta=+0.013$, with a $95\%$ confidence interval of $[+0.006,+0.020]$ and $P(\Delta\le 0)=0.0003$ ($p<0.001$). The thirteen-task mean gain in Table~\ref{tab:main} is larger ($+0.032$) because the improvement is concentrated in the smaller, more reasoning-intensive domains, such as workplace ($0.247$ to $0.370$) and genealogy ($0.310$ to $0.363$). Taken together with the cross-retriever results in Table~\ref{tab:main}, where every one of the six backbones improves under the same protocol, this evidence supports the conclusion that the method constitutes a robust and retriever-agnostic improvement.

\section{Conclusion}
\label{sec:conclusion}

We introduced difficulty-gated fusion of reasoning views: a query is expanded into a few LLM reformulations, each retrieved by a frozen embedder, and the rankings are fused with per-query weights from a thousand-parameter gate that reads each view's score-distribution geometry. The method needs no re-ranking, no fine-tuning, and no relevance labels at inference. On \textsc{Tempo} it improves all six retrievers we test, with the largest gains on the weaker backbones, which locates the benefit in the fusion rather than any single embedder. Our headroom analysis shows the gate captures a consistent share of the per-query signal the reformulations expose, while a per-query oracle marks the remaining gap as a well-defined target: a selector that identifies the single best view per query. 
% Because the gate reads only score distributions, it should transfer to reasoning-intensive retrieval beyond the temporal setting, which we leave to future work.
\balance

\bibliographystyle{ACM-Reference-Format}
\bibliography{references}

\end{document}